\documentclass[usenatbib]{mn2e}
\usepackage{txfonts}
\usepackage{graphicx}
\usepackage{array}
\usepackage{longtable}
\bibpunct{(}{)}{;}{a}{}{,}

\title[Open cluster NGC 7142] {The open cluster NGC 7142: interstellar
extinction, distance and age}

\author[V. Strai\v{z}ys et al.]
{ V. Strai\v{z}ys,$^{1}$\thanks{E-mail:vytautas.straizys@tfai.vu.lt} M.
Maskoli\={u}nas,$^{1}$ R. P. Boyle,$^{2}$ K. Zdanavi\v{c}ius,$^{1}$ J.
Zdanavi\v{c}ius,$^{1}$ \newauthor V. Laugalys$^{1}$ and A.
Kazlauskas$^{1}$\\
\\
     $^{1}$~Institute of Theoretical Physics and Astronomy, Vilnius
University, Go\v{s}tauto 12, Vilnius 01108, Lithuania\\
    $^{2}$~Vatican Observatory Research Group, Steward Observatory,
 Tucson, AZ 85721, U.S.A. }

\begin{document}

\date{Accepted 2013 October 16. Received 2013 October 15; in original
form 2013 August 5}
\pagerange{1--8} \pubyear{2013}

\maketitle

\label{firstpage}

\begin{abstract} The results of medium-band photometry of 1037 stars in
the area of old open cluster NGC 7142 down to $V$ = 20.1 mag in the
Vilnius seven-colour system are presented.  Photometric results are used
to classify in spectral and luminosity classes about 80\,\% of stars
down to $V$ = 18.5 mag, to identify cluster members, to determine the
main cluster parameters and to investigate the interstellar extinction
in this direction.  The average extinction $A_V$ of the cluster is about
1.1 mag ($E_{B-V}$ = 0.35), and its distance is 2.3 kpc (the distance
modulus 11.8 mag).  The age of the cluster, 3.0 Gyr, is estimated from
the intrinsic colour-magnitude diagram with individual dereddening of
each star and the Padova isochrones.  The surface distribution of the
extinction is shown.  The reddening of the eclipsing variable V375 Cep
is found to be close to the average reddening of the cluster.  Probably,
the cluster contains five red clump giants, two asymptotic branch stars
and four blue stragglers.  \end{abstract}

\begin{keywords} stars:  fundamental parameters -- ISM:  extinction --
Galaxy:  open clusters and associations:  individual:  NGC 7142
\end{keywords}

\section{Introduction}

The open cluster NGC 7142 is an important object, being one of the old
clusters which is used in plotting various relations between structural,
chemical and evolutionary parameters of the cluster system
\citep{Phelps1994, Janes1994, Friel1995, Twarog1997, Carraro1998,
Salaris2004, Cheng2012}.  Every improvement of the accuracy of
parameters of these clusters is an important input to understanding the
structure and evolution of the Galactic disk.  However, due to uneven
distribution of interstellar reddening and considerable contamination by
field stars, the parameters of NGC 7142 are still far from being of
sufficient accuracy.

The cluster is located at the Galactic latitude +9.5$\degr$, close to
the boundary of the Cepheus Flare, a branch of the Local spiral arm
deviating towards the North Celestial Pole.  The surrounding area
exhibits a complicated pattern of molecular and dust clouds, see the
recent review of this area by \citet{Kun2008}.  At only 0.4$\degr$ from
NGC 7142, a well-known reflection nebula NGC 7129 with the embedded
young cluster is located.  The nebula is a part of the clump P2 of the
dust cloud TGU H645 identified in the \citet{Dobashi2005} atlas of dark
clouds.  The distribution of the 100 $\mu$m emission
\citep{Schlegel1998} shows that NGC 7142 can be partly covered by the
periphery of the same dust cloud.

The first photometric investigations of NGC 7142 in the {\it UBV} system
published in the 1960s, have used the photographic method with standard
stars measured photoelectrically.  The first $V$ vs.  $B$--$V$ diagram
of NGC 7142 was published by \citet{Hoag1961}, which exhibited the
sequence of red giants and a crowding of stars near the turn-off point
of the main sequence.  In the presence of large scatter, the limiting
magnitude ($V$ = 16.5) was not sufficient to cover main sequence stars
below the turn-off point.  Therefore, the reddening and distance of the
cluster were estimated only approximately \citep{Johnson1961}.
Analyzing the colour-magnitude diagram (hereafter CMD) of NGC 7142,
\citet{vandenBergh1962} first noted that its general features show a
strong resemblance to those of the old open clusters M67 and NGC 188.
He also directed attention to the presence of irregularities of the
absorbing cloud close to the cluster.  \citet{Sharov1968} described a
group of stars close to the main sequence but located to the left from
the turn-off point (blue stragglers).  The next photometric
investigation of NGC 7142 was published by \citet{vandenBergh1970} down
to limiting $V$ = 17 mag, and this allowed to estimate more reliable
colour excess and age of the cluster.

The first CCD photometry of the cluster in the {\it BV} system down to
$V$ = 18 mag was published by \citet{Crinklaw1991}.  They confirmed a
considerably variable extinction across the face of the cluster
suspected earlier by other authors -- differential reddening was found
to be of the order of $\Delta E_{B-V}$ = 0.1.  Their CMD was dereddened
with a mean value of $E_{B-V}$ = 0.35 and gave the distance modulus 11.4
mag by fitting to the zero-age main sequence (ZAMS).  They also noted
that a binary star population is present in the cluster.

The most recent CCD photometry of NGC 7142 has been done by
\citet{Janes2011} and \citet{Sandquist2011} in the {\it BVI} system,
down to $V$ = 20--21 mag.  Both investigations for determining the
distance and age have used the red clump giants (hereafter RCGs) and the
turn-off point of the main sequence, comparing their positions with
isochrones and with the old cluster M67.  In both papers the resulting
colour excesses are in full agreement, $E_{B-V}$ = 0.32. The true
distance moduli, $(m-M)_0$, are also very close, 11.85 and 11.9 mag.
However, the age of the cluster determined by Janes et al. is 6.9 Gyr,
while Sandquist et al. find $\sim$\,3 Gyr.

Probably, there are several reasons for the differences in age discussed
by Sandquist et al.  The main difficulty in deriving cluster properties
is related to contamination of the CMD with field stars, and to the
reddening and extinction differences across the cluster field.  The last
effect makes all branches of NGC 7142 very broad and prevents to make
reliable shift of isochrones for the age determination.  To make the
sequences narrower, individual determination of reddening for each
cluster star is needed.  Memberships to the cluster from radial
velocities are known only for 16 red giants \citep{Friel1989, Friel1993,
Jacobson2007, Jacobson2008, Sandquist2011}.

For determining the extinction for individual stars we need their
two-dimensional classification which can be done either by
spectroscopy or by multicolour photometry.  The Vilnius seven-colour
system with the mean wavelengths at 345, 374, 405, 466, 516, 544 and 656
nm was developed especially for a similar aim.\footnote{~For the
description of the Vilnius photometric system see \citet{Straizys1992}
monograph available in pdf format at\\
http://www.itpa.lt/MulticolorStellarPhotometry/ and ADS.}

About a hundred stars of NGC 7142 using CCD photometry in the Vilnius
system were classified in MK types in our earlier paper
\citep{Maskoliunas2012}.  That investigation included 2140 stars down to
$V$ = 17 mag in a 1.5 sq. degree field with the NGC 7129 and NGC 7142
clusters and a broad surrounding area.  For about 60\% of stars
photometric two-dimensional classification has been given.  However, the
limiting magnitude of that catalogue was not sufficient for the
investigation of the NGC 7142 cluster.  Therefore we decided to perform
new CCD photometry of the cluster with a telescope having a better
resolution and a deeper limiting magnitude.

\section{Observations, their processing and the classification of stars}

The observations were obtained by one of the authors (R.  P. Boyle) in
2009 October 18--21 with the 1.8 m telescope of the Vatican Observatory
on Mt. Graham (Arizona) equipped with a 4K backside illuminated CCD
camera and liquid nitrogen cooling.  The camera contains a
62\,$\times$\,62 mm chip which gives a 13\arcmin\,$\times$\,13\arcmin\
field-of-view, with a scale of 0.38\arcsec/pixel (binned
2\,$\times$\,2).  The cluster was framed with a set of exposures (Table
1) to ensure the linearity of response from $V$\,$\approx$\,9 mag to the
faintest limit.  The seeing during the observations usually was of the
order of 1$\arcsec$.  The field center is at (J2000):  21$^{\rm
h}$45$^{\rm m}$10$^{\rm s}$, +65$\degr$\,46$\arcmin$\,30$\arcsec$.

\begin{table}
\caption{Log of CCD observations. The columns give the
filter name, the mean wavelengths of the standard system, the range
of exposure lengths and the total number of frames in each filter.}
\label{table:1}
\centering
\begin{tabular}{cclc}
\hline\hline
 Filter & $\lambda_0$ nm & Exposure range & Number of frames.
\\
\hline
\noalign{\vskip0.5mm}
$U$ & 345 & From  250 s to 25 s  &  15  \\
$P$ & 374 & From  800 s to 30 s  &  12  \\
$X$ & 405 & From  500 s to 25 s  &  ~8  \\
$Y$ & 466 & From  120 s to  4 s  &  12  \\
$Z$ & 516 & From  300 s to  4 s  &  12  \\
$V$ & 544 & From  120 s to  4 s  &  10  \\
$S$ & 656 & From  240 s to  8 s  &  14  \\
\hline
\end{tabular}
\end{table}


\begin{figure}
\resizebox{\hsize}{!}{\includegraphics{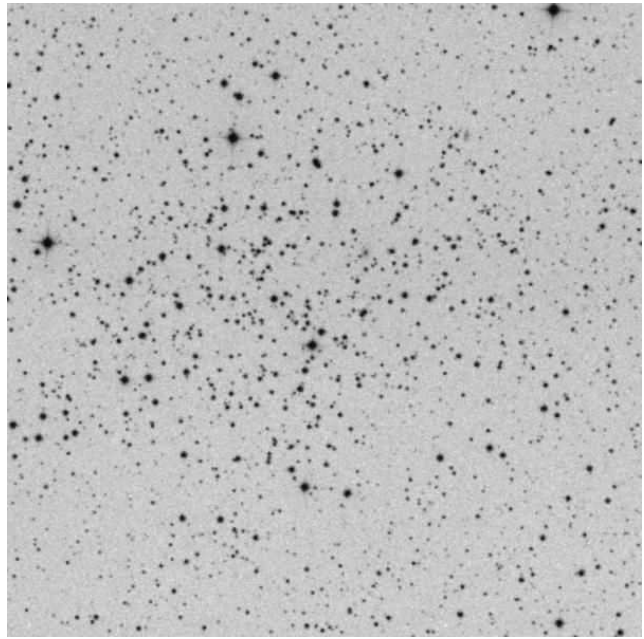}}
\vskip1mm
\caption{Area of the cluster NGC 7142 observed in
the Vilnius photometric system with
VATT (13$\arcmin$\,$\times$\,13$\arcmin$).
The DSS2 Red image from SkyView.}
\label{1}
\end{figure}


\begin{figure}
\resizebox{\hsize}{!}{\includegraphics{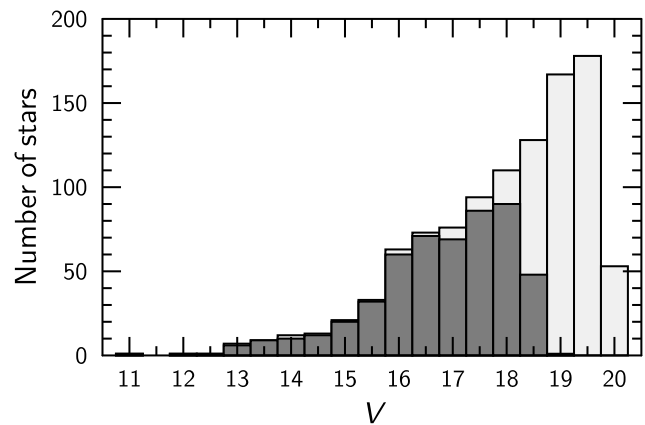}}
\vskip1mm
 \caption{Distribution of the measured stars in the NGC 7142 area
in apparent magnitudes. The shadowed parts of the columns correspond to
stars for which two-dimensional spectral types were determined.}
\label{2}
\end{figure}

For the reduction of CCD exposures, the IRAF program package in the
aperture photometry mode was used.  The radii of apertures were
1.5--2.0\arcsec, depending on the star image sizes. For flat-fielding,
the twilight and dome exposures were applied.  Large-scale systematic
errors in the flat fields were corrected by using the exposures of the
cluster M67 with known photometric data in the Vilnius system of high
accuracy \citep{Laugalys2004}.  Preliminary colour equations for the
reduction of magnitudes and colour indices of stars from the
instrumental to the standard system were obtained also from observations
of M67. Zero-points of magnitudes in the Vilnius system were based on
117 common stars with the \citet{Maskoliunas2012} catalogue.  The final
adjustment of colour equations and zero-points has been done by
optimizing the accuracy of photometric classification of a selected set
of standard stars in the investigated area (V.  Laugalys, in
preparation).

Magnitudes and colour indices were determined for 1037 stars down to the
limiting magnitude 20.1.  The distribution of stars in magnitudes is
shown in Fig.\,2.  The shadowed parts of the columns correspond to stars
for which spectral and luminosity classes in the MK system were
determined from the Vilnius photometric data.  They compose 80\,\% down
to $V$ = 18.5 mag.  The accuracy of the magnitudes $V$ and colour
indices $X$--$V$, $Y$--$V$, $Z$--$V$ and $V$--$S$ down to $V$ = 16 mag
is usually better than 0.02 mag, the errors of $U$--$V$ and $P$--$V$ are
about 1.5--2.0 times larger.

The classification of stars was based on two methods.  One of them uses
a set of 14 interstellar reddening-free $Q$-parameters compared with the
corresponding parameters for about 8300 standard stars with known MK
types \citep{Straizys2013}.  The second method uses 300 synthetic
standard stars constructed from the mean intrinsic colour indices for
different MK types and reddened with the interstellar extinction law (K.
Zdanavi\v{c}ius, in preparation).  In the classification, the normal
interstellar extinction law was accepted.  Normality of the law in the
dust cloud TGU H645 has been verified by determining the ratio of colour
excesses $E_{J-H} / E_{H-K_s}$ for red giants with the data from the
2MASS survey.

The accuracy of spectral types is characterised by the values of $\sigma
Q$, the match quality of $Q$ parameters between the program star
and standard stars:
\begin{equation}
\sigma Q = \pm\sqrt{{\sum_{n}^{} \Delta Q_i^2}\over n},
\end{equation}
where $\Delta Q$ are differences of corresponding $Q$-parameters of the
program star and a standard star, $n$ is a number of the compared
$Q$-parameters (in our case $n$ was between 10 and 14).
 If the $\sigma Q$ is $\leq 0.02$ mag, the
classification is of good accuracy, if it is within 0.02--0.03 mag, the
accuracy is reasonable.  The quality of spectral types is low if $\sigma
Q$\,$>$\,0.03 mag.  However, for K-type stars $\sigma Q$s up to 0.04 mag
are acceptable.  K-stars can be classified in spectral and luminosity
classes even without the ultraviolet color indices $U$--$V$ and
$P$--$V$.  A typical 3$\sigma$ classification errors in spectral classes
are $\pm$\,2 decimal subclasses for A-F-G stars and $\pm$\,0.5 subclass
for K stars.  The errors in luminosity classes are $\pm$\,1 class for
A-F-G stars and $\pm$\,0.5 class for K stars.  At $V$\,$>$\,18.5 mag the
accuracy of photometry is too low for reliable classification of stars.
For most of these stars the $U$-$V$ and $P$--$V$ colour indices are not
given.

The main problem in photometric classification are the stars with
various peculiarities and unresolved binaries.  The suspected
metal-deficient stars (designated `md') require spectroscopic
confirmation.  In the present investigation we do not use the stars
classified as peculiar and with low accuracy ($\sigma Q$\,$>$\,0.03 mag,
except of K-stars).

\section{Catalogue}

Magnitudes and colours for stars in the NGC 7142 area are listed in
Table 2. The printed text contains only a sample of the table with 10
stars.  The sample is taken from the central part of the catalogue to
represent all the columns.  The full table is accessible only in the
online version.

The columns list the following information:  star number, equatorial
coordinates J2000.0, magnitude $V$, colour indices \hbox{$U$--$V$},
$P$--$V$, $X$--$V$, $Y$--$V$, $Z$--$V$ and $V$--$S$, photometric
spectral type in the MK system, accuracy of spectral types $\sigma Q$,
interstellar extinction $A_V$, the possible membership to the cluster
(see the next section) and the note number.  The colour indices with
$\sigma$ = 0.05--0.10 mag are marked with colons.  Spectral classes are
designated in the lower-case letters to indicate that these are
determined from photometric data.  Notes at the end of the table contain
important information on the stars available in the literature.  The
stars found to be binaries or having asymmetrical images were not
classified -- they are designated with the numeral 4 in the last column.
The coordinates of stars were taken from the PPMXL catalogue
\citep{Roeser2010} with rounding to two decimals of time second and one
decimal of arcsecond.


\begin{table*}
\begin{minipage}{190mm}
\def\hstrut{\vrule height10pt depth0pt width0pt}
\def\lstrut{\vrule height0pt depth6pt width0pt}
\caption{A sample of the photometric catalogue in the NGC 7142 area
for stars measured in the Vilnius seven-colour system.}
\label{table2}
\begin{tabular}{rrlllllllllcccc}
\hline\hline
 No. &  RA\,(J2000) & DEC\,(J2000) & $V$~~~~ & $U$--$V$ &
$P$--$V$ &  $X$--$V$ & $Y$--$V$ & $Z$--$V$ & $V$--$S$ & Phot. & $\sigma
Q$ & $A_V$ & Member- & Notes \hstrut \\
&  h~~~m~~~s~~~ & ~~$\circ$~~~$\prime$~~~$\prime\prime$ &
mag~~ & mag  & mag & mag & mag & mag & mag  & sp. type & mag & mag &
ship & \lstrut \\
\hline
\noalign{\vskip 0.5mm}
  681   &21 45 20.28 & +65 48 10.2 &  16.836 &  2.933  &  2.322  &  1.665  &  0.776  &  0.272  &  0.780  &   f7\,V    &  0.015 &  1.21 & m &     \\
  682   &21 45 20.34 & +65 41 30.9 &  18.513 &  3.002: &  2.626  &  1.827  &  0.737  &  0.269  &  0.841  &            &        &       &   &     \\
  683   &21 45 20.43 & +65 48 31.0 &  13.779 &  4.171  &  3.447  &  2.449  &  1.031  &  0.381  &  0.982  &   g8\,III  &  0.017 &  1.29 & m &  1,2,3\\
  684   &21 45 20.46 & +65 51 07.0 &  12.982 &  2.883  &  2.304  &  1.586  &  0.698  &  0.262  &  0.620  &   f9\,IV-V &  0.009 &  0.79 &   &  1  \\
  685   &21 45 20.88 & +65 48 41.5 &  17.395 &  3.101  &  2.551  &  1.817  &  0.829  &  0.308  &  0.802  &   g0\,V    &  0.010 &  1.25 & m &     \\
  686   &21 45 20.89 & +65 47 39.9 &  12.792 &         &         &  2.705  &  1.082  &  0.415  &  0.923  &   g5\,II   &  0.007 &  1.25 &   &  1,2\\
  687   &21 45 20.91 & +65 46 39.5 &  15.528 &  2.955  &  2.308  &  1.625  &  0.751  &  0.249  &  0.732  &   f4\,V    &  0.019 &  1.29 &   &     \\
  688   &21 45 20.94 & +65 49 13.6 &  18.471 &  2.815  &  2.226  &  1.629  &  0.754  &  0.274  &  0.753  &  f-g,\,md: &        &       &   &     \\
  689   &21 45 21.08 & +65 47 02.8 &  16.619 &  3.018  &  2.422  &  1.737  &  0.781  &  0.290  &  0.730  &   f9\,V    &  0.005 &  1.08 & m &     \\
  690   &21 45 21.10 & +65 42 02.2 &  14.020 &  3.591  &  3.044  &  2.070  &  0.844  &  0.342  &  0.814  &   k0\,IV   &  0.018 &  0.58 &   &     \\
\hline
\end{tabular}
\vskip1mm
{\bf Notes.}\\
The columns give the running numbers, equatorial coordinates, $V$
magnitude, colour indices, photometric spectral type and its accuracy,
interstellar\\ extinction, cluster membership and notes.  The full
catalogue of 1037 stars is available online.
\end{minipage}
\end{table*}

\section{Cluster distance}

Interstellar extinctions and distances of stars were calculated
with the equations
\begin{equation}
A_V = 4.16~[(Y-V)_{\rm obs} - (Y-V)_0],
\end{equation}
\begin{equation}
5 \log d = V - M_V +5 - A_V,
\end{equation}
where the intrinsic colour indices $(Y-V)_0$ and absolute magnitudes
$M_V$ for the field stars were taken from the calibrations of spectral
and luminosity classes \citep{Straizys1992}.  The extinctions $A_V$
in the medium-band Vilnius system and in the broad-band {\it UBV} system
are very close. For the cluster members, the absolute magnitudes were
taken individually as will be described lower.  Thus, our first task was
an attempt to identify possible cluster members.

First, we calculated  the extinction-corrected magnitudes
\begin{equation}
V_0 = V_{\rm obs} - A_V
\end{equation}
for all stars with two-dimensional classification and plotted the
intrinsic $V_0$ vs.  $(Y-V)_0$ diagram (hereafter, intrinsic CMD), where
the intrinsic colour indices were taken according to MK spectral
types.

As was mentioned in the Introduction, for some stars in the field,
mostly for red giants, membership to the cluster is known from their
radial velocities.  Among them, most important are RCGs, which can be
applied to estimate the distance to the cluster.  Then the ZAMS line and
the isochrones for the known distance modulus can be plotted on the
intrinsic CMD, and this allows to locate the position of the cluster's
main sequence.

The next step is the exclusion of majority of the foreground and
background stars.  Preliminary distances for all stars can be calculated
with the $M_V$ values taken according to their MK types.  These
distances for the cluster stars, due to the errors of absolute
magnitudes and the presence of unresolved binaries, should exhibit a
considerable scatter.  In the photometric classification of A-F-G stars
we usually accept $\pm$\,0.5 mag as the 3$\sigma$ error of $M_V$.  In
this case the cluster stars with the maximum distance errors should
appear less or more distant by a factor of 1.26.  Thus, if the true
distance of the cluster is $d$, we should expect that the cluster stars
due to luminosity errors will be scattered between distances $d$/1.26
and $d$\,$\times$\,1.26.  The stars with lower and higher distances
can be considered to be foreground and background stars.  They may
be excluded from the dereddened CMD of the cluster.

Another source of scatter of the cluster stars to smaller distances are
unresolved binaries.  If both components of a binary star are of the
same luminosity, the distance of such system is reduced by a factor
of 1.41.

For determining the distance to NGC 7142 we applied five RCGs identified
by \citet{Sandquist2011}.  The numbers of these stars in Table 2 are
461, 493, 683, 998 and 1026, all of them are of spectral types G8--G9
III classified with good accuracy.  The sixth RCG identified by
Sandquist et al., our No.\,799, could not be classified since it is a
visual binary with the components of similar brightness and a separation
of 2--3$\arcsec$.  The mean distance to the five RCGs, 2.33\,$\pm$\,0.14
kpc, was calculated taking their average extinction-corrected magnitude
$V_0$ = 12.54 mag and the absolute magnitude $M_V$ = +0.7 mag,
corresponding to G8--G9 giants in the RCG sequence on the $M_V$ vs.
$B$--$V$ diagram for the Hipparcos stars \citep{Perryman1997}.  The
given rms error originates from the dispersion of extinction-corrected
$V$ magnitudes of the five stars.

Distances to the same five RCGs were also calculated from 2MASS {\it
JHK}$_s$ photometry \citep{Skrutskie2006} accepting their absolute
magnitude $M_{K_s}$ = --1.6 \citep{Alves2000, Grocholski2002} and the
intrinsic colour index $H$--$K_s$ = 0.09
\citep{StraizysLazauskaite2009}.  The average distance of these stars,
2.27\,$\pm$\,0.18 kpc, is in good agreement with the result from Vilnius
photometry.  For the further analysis of the cluster, we accept it is
located at a distance of 2.30\,$\pm$\,0.16 kpc, corresponding to the
true distance modulus $V$--$M_V$ = 11.8 mag.  Due to the errors of
absolute magnitudes ($\pm$\,0.5 mag), cluster members at this distance
should be scattered between 1.8 and 2.9 kpc.

Except of the five RCGs, we also attribute to cluster members a few red
giants of spectral class K for which the membership was inferred from
radial velocities, references are given in the Introduction.  The three
reddest giants of spectral types K2--K4.5 with numbers 1077, 1102 and
1181 were taken from the catalogue of \citet{Maskoliunas2012}.  A few
G-K giants and subgiants were added to the list of members according to
their position in the dereddened CMD and the distance, with possible
errors of $M_V$ taken into account.

\section{Cluster stars near the turn-off point}

In the intrinsic CMD (Fig.\,4, see lower) a strong crowding of stars is
seen near the main sequence with $(Y-V)_0$ from 0.46 to 0.53,
corresponding to the spectral classes F5--G0 and the luminosity classes
V--IV.  This signifies the presence here of the cluster stars near the
turn-off point.  We have accepted that all these crowding stars are
cluster members since the main sequence in the direction of higher
temperatures is almost empty (except of a few possible blue stragglers).

Unfortunately, in this spectral range the luminosity effect between V
and IV classes on $Q$-parameters is quite small, and most stars located
close to the turn-off point during photometric classification have been
attributed to luminosity V. However, in the intrinsic CMD many of them
lie at different levels above ZAMS, consequently, they should be of a
higher luminosity.  To make the absolute magnitudes closer to reality,
we took into account their rise in the CMD diagram above the ZAMS line,
$\Delta V$, which was subtracted from $M_V$ of the ZAMS.  This procedure
was not applied only to the stars located $\pm$\,0.2 mag from the ZAMS
-- these stars were accepted to belong to luminosity V. In this way more
negative absolute magnitudes were found for many stars in the region of
the turn-off point.  Since some stars in this region of CMD are
binaries, their luminosities determined by this method should be
overestimated.  This must be taken into account determining the age of
the cluster.

The choice of the coolest stars at spectral class G0, to which the
absolute magnitudes have been corrected with the described method, is
somewhat voluntary. It is based on the fact that for G-type
stars the photometric luminosity effect is larger, and the stars
are classified in luminosity classes with a better reliability.

\section{Intrinsic colour-magnitude diagram}


\begin{figure}
\resizebox{\hsize}{!}{\includegraphics{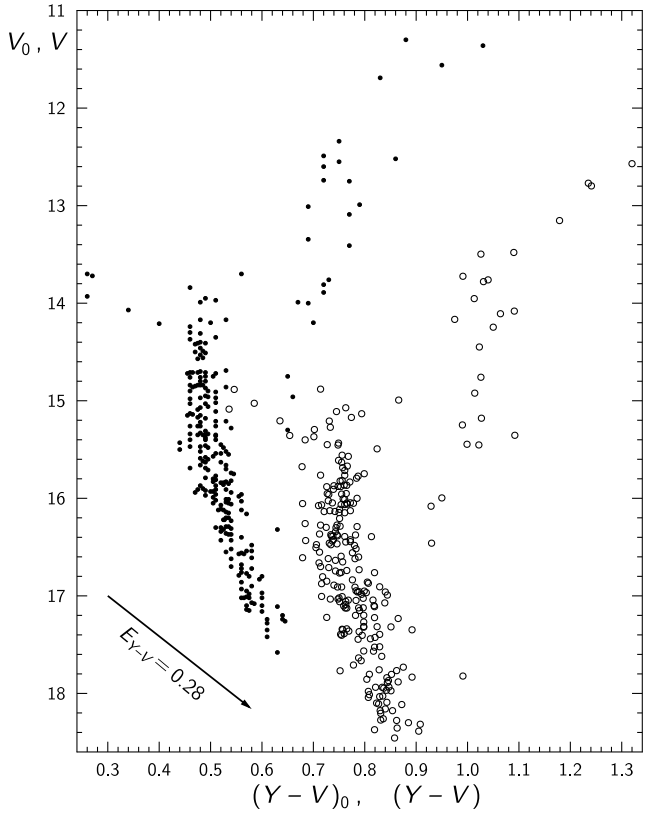}}
\caption{Colour-magnitude diagram for 255 possible members
NGC 7142.  Open circles designate the observed positions of
stars, and  dots are the same stars with the individually dereddened
colours and the extinction-corrected magnitudes. The arrow is the
reddening line for $E_{Y-V}$ = 0.28, the average value of the cluster
reddening.}
\label{3}
\end{figure}


\begin{figure}
\resizebox{\hsize}{!}{\includegraphics{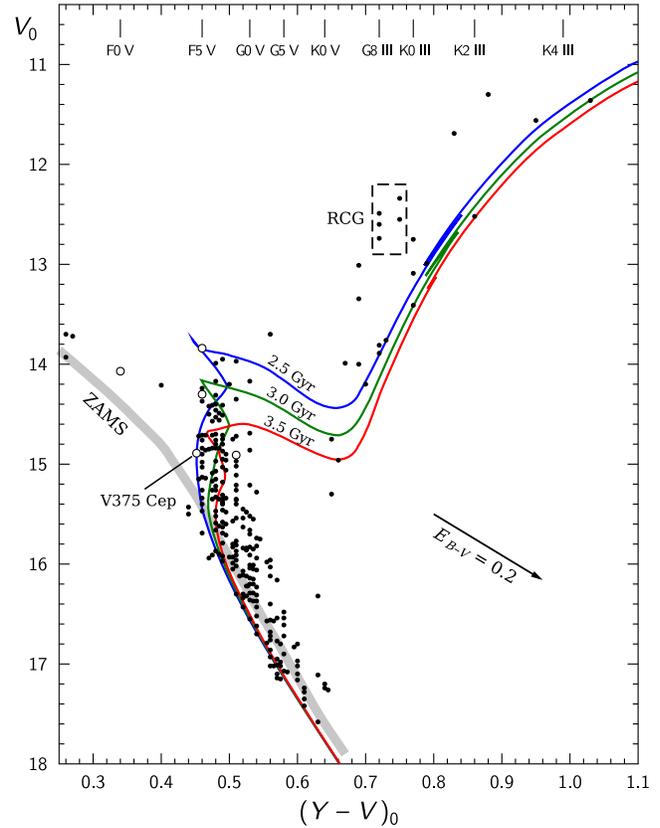}}
\caption{Intrinsic colour-magnitude diagram for 255 possible members of
NGC 7142.  The grey line is the ZAMS from
\citet{Kazlauskas2006} and the three coloured lines are the Padova
isochrones for the ages 2.5, 3.0 and 3.5 Gyr, all are shifted according
to a distance modulus of $V_0$--$M_V$ = 11.8.  Spectral classes,
corresponding to the intrinsic $(Y-V)_0$ colours, are indicated at the
top.  The five possible RCG stars are placed in a rectangular box.
Some amount of field stars can be present in the diagram. Five
eclipsing variables are plotted as open circles. One of them,
V375 Cep, is the cluster member. To make all stars
visible, the overlapping dots were shifted along the $(Y-V)_0$ axis by
+0.005 or --0.005 mag.}
\label{4}
\end{figure}

Fig.\,3 shows the observed and the intrinsic CMDs for 255
possible cluster members which include: (1) all stars in the range of
distances 1.8--2.9 kpc, (2) stars of spectral classes F5-G0 and
luminosity classes V--IV, (3) red giants with the membership inferred
from radial velocities, and (4) a few red giants and subgiants whose
spectral types and distances are in reasonable agreement with the
cluster membership.  In Table 2 the suspected cluster members are
indicated by `m' in the next-to-last column.

In Fig.\,3 circles show the observed positions of stars and dots
are the same stars shifted along the reddening lines according to their
$E_{Y-V}$ and $A_V$ values.  The shown length of the reddening line
corresponds to $E_{B-V}$ = 0.35, the mean value of color excess of the
cluster.  It is evident that individual dereddening resulted in a
considerable decrease of scatter of stars, especially on the main
sequence.

Fig.\,4 shows the intrinsic CMD of NGC 7142, together with
isochrones, ZAMS line and other information.  The CMD can be
contaminated by field stars -- some of them can be mixed together with
the cluster members in sequences, others are easily seen by their
outlying positions.  For example, the circles show five eclipsing
variables four of which are non-members \citep{Sandquist2011,
sandquist2013}.  The accuracy of positions for the main sequence stars
and K-giants mainly depends on the errors of spectral classes.  Near the
main sequence these errors are of the order of two subclasses, and this
results in $\Delta (Y-V)_0 \approx$\,0.03 mag.  The position errors of
K-giants are similar, because larger changes in colour between K
subclasses are compensated by a larger accuracy of spectral classes
($\pm$\,0.5 of the subclass).  The errors in magnitudes and colours are
correlated:  the spectral class errors move the dereddened star along
its reddening line with the slope $A_V$/$E_{Y-V}$ = 4.16.

Because the intrinsic colour indices $(Y-V)_0$ were taken according to
the MK spectral types, the stars have a discrete distribution along
the colour axis.  Thus, the appearance of CMD of NGC 7142 is somewhat
different from those which are plotted directly from observational data
and dereddened using a fixed value of the colour excess.  The width of
reddening-free cluster sequences in the CMD of NGC 7142, produced by our
method, appears to be wider than the similar sequences of an old open
cluster with low reddening, such as M67 \citep{Boyle1998}.

The grey line in Fig.\,4 represents ZAMS for solar metallicity from
\citet{Kazlauskas2006} corresponding to the distance modulus
$V_0$--$M_V$ = 11.8 mag.  The main sequence of the cluster has the
turn-off point at ($Y$--$V$)$_0$ = 0.46, spectral type F5\,V.  The
spread of stars in absolute magnitudes at this point is about 1.5 mag.
About 15 stars are located close to the expected sequence of giants and
subgiants.  Four possible blue stragglers (their numbers in Table
2:  337, 418, 664 and 754, spectral classes A6--F2) are seen close to
the ZAMS.  Due to possible duplicity, their positions in the
colour-magnitude diagram can be of lower accuracy.  One of the
stars (No.\,631, F0\,V), looking as a blue straggler, is an eclipsing
variable V3 from the \citet{Sandquist2011} list, non-member of the
cluster. Two K-stars, Nos. 348 (K2.5\,III) and 820 (K1.5\,III) can
belong to the asymptotic giant branch.

The intrinsic CMD of the cluster can be used for its age determination
comparing to isochrones, but for this the metallicity should be known.
From photometric diagrams in the Washington system, [Fe/H] = --0.17 was
found by \citet{Geisler1991}.  From the medium-resolution spectroscopy
of 11 red giants \citet{Friel1993} found the mean metallicity [M/H] =
0.0 from the Fe and Fe-peak element blends and --0.23 from the Mgb+MgH
feature.  \citet{Twarog1997} transformed these values to a revised
metallicity scale and received a value of [Fe/H] = +0.04.
\citet{Jacobson2007, Jacobson2008} from medium- and high-resolution
spectra of six stars have found [Fe/H] = +0.08 and +0.14.  Sandquist et
al.  (2013) from high-resolution spectra of the eclipsing variable V375
Cep have found [Fe/H] = +0.09.  Thus for the age determination we
decided to use the isochrones for [Fe/H] = +0.10. If the solar
metallicity is $Z$ = 0.014 \citep{Asplund2009} or 0.015
\citep{Caffau2011}, [Fe/H] = +0.10 corresponds to $Z$ =
0.018--0.019.

In Fig.\,4 we plot three isochrones for the ages 2.5, 3.0 and 3.5 Gyr
from the Padova database of stellar evolutionary tracks and
isochrones\footnote{\citet{Bressan2012} and
http://stev.oapd.inaf.it/cgi-bin/cmd} for the Vilnius system adjusted to
a distance modulus of 11.8 mag.  It is evident that the main-sequence
lines of the isochrones with respect to the observed ZAMS line show a
shift down by 0.2 mag in absolute magnitudes {\it or} by 0.02 mag to the
left in $Y$--$V$.  It is difficult to determine the reason of this
shift; it can be the result both of the observational ZAMS and the
theoretical isochrones.  The problems of the transformation of
isochrones to the observational plane were discussed in many papers, see
e.g.  \citet{VandenBerg2010} and \citet{Sandquist2011}.

A visual comparison of the distribution of stars near the turn-off point
with isochrones shows that the cluster age should be somewhere between
2.5 and 3.0 Gyr.  In the region of red giants, the 3.0 Gyr isochrone
seems to be preferable.  However, if we shift the isochrones by 0.2 mag
upward (up to the coincidence with the observed ZAMS), then the 3.0 and
3.5 Gyr isochrones become favoured.  Also, if the four stars at the hook
of the upper isochrone between $V_0$ = 13.8 and 14.0 mag are binaries or
non-members, the 3.0 Gyr isochrone would well represent its hook stars
at 14.2--14.4 mag.  The bluest star on the 2.5 Gyr isochrone, No.\,239
(V1,
F5\,V), is really the binary, eclipsing variable and probable non-member
(Sandquist et al. 2011, 2013).  Consequently, there are good reasons to
consider that the age of NGC 7142 is close to 3.0\,$\pm$\,0.5 Gyr.

\section{Extinction vs. distance and its surface distribution}

The $A_V$ vs.  $d$ diagram for 507 stars in the direction of NGC 7142,
derived from the data of Table 2, is shown in Fig.\,5.  It includes both
field stars and the cluster members.  At low distances the extinction
$A_V$ increases gradually up to 1.0--1.2 kpc reaching the values between
0.8 and 1.5 mag.  At this distance range the line of sight should cross
the outskirts of the dust cloud TGU H645 P2 in which the nearby young
cluster NGC 7129 is embedded.  At larger distances no extinction rise is
expected since at this Galactic latitude our line of sight recedes from
the Galactic dust layer near its plane.  At a distance of 1.2 kpc the
line of sight is already 200 pc above the plane.


\begin{figure}
\resizebox{\hsize}{!}{\includegraphics{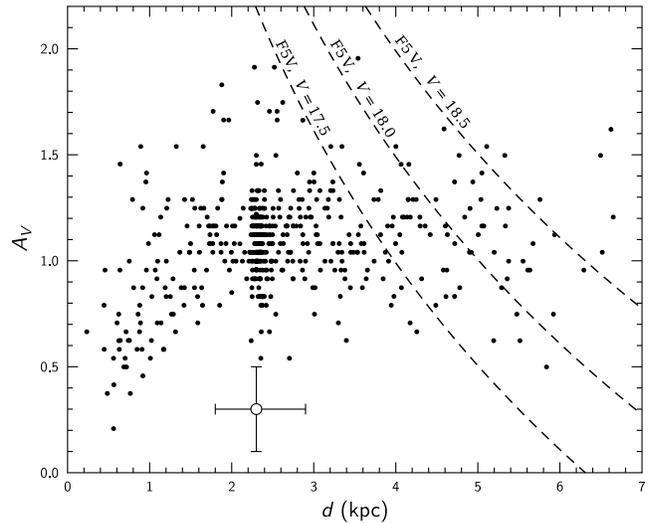}}
\caption{Extinction vs. distance diagram for 507 stars of
the NGC 7142 area classified in spectral and luminosity classes.
The open circle with error bars is located at the cluster distance.
The three curves show the limiting magnitude effect for
F5\,V stars of the apparent magnitudes 17.5, 18.0 and 18.5.}
\label{5}
\end{figure}

The vertical bar of stars in Fig.\,5 at $d$ = 2.3 kpc and $A_V$
between 0.8 and 1.35 mag is formed by the cluster members, mostly by
F5-G0 stars for which the absolute magnitudes were calculated as
described in Section 5.

In the area, the maximum number of stars with two-dimensional spectral
types falls on 17--18 mag.  Most of them are main-sequence stars of
spectral classes F and G. The three broken curves in Fig.\,5 demonstrate
the effect of limiting magnitude for F5\,V stars with magnitudes $V$
between 17.5 and 18.5.  Above the upper curve, only B- and A-type stars,
as well as G-K-M giants can be found.  These types of stars are rare in
this area.

{In Fig.\,5 we show the surface distribution of the extinction over the
area.  The 466 stars with the distances above 1.15 kpc (the distance to
the dust cloud TGU H645) have been divided into five ranges of the
extinction and plotted on the plane of equatorial coordinates as circles
of different sizes.  The most prominent feature of this map is the
filamentary distribution of stars with almost empty spaces between them.
This feature is also well seen in CCD exposures with different filters.
Also, the stars in some of these chains and groups exhibit similarity of
extinctions.  However, it is too problematic to locate large-scale
boundaries between subareas with similar values of extinction since the
picture is too mottled.


\begin{figure}
\resizebox{\hsize}{!}{\includegraphics{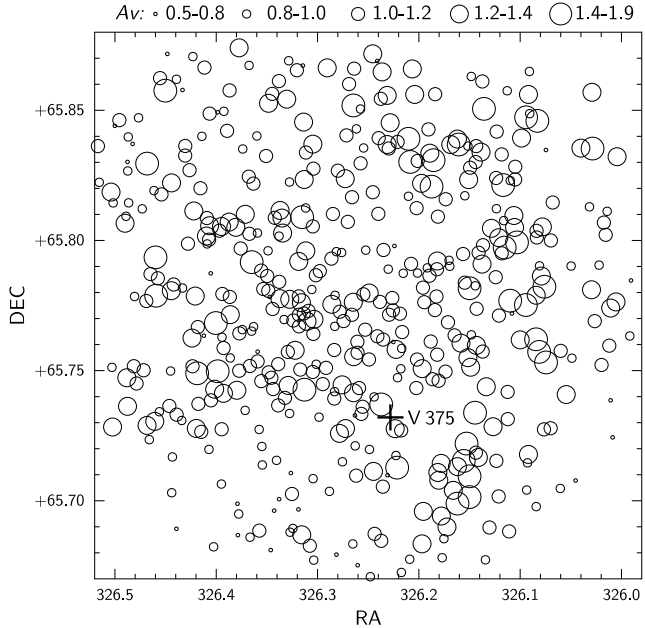}}
\caption{Surface distribution of the extinction for 466 stars
with distances $>$\,1.15 kpc (a distance of the dust cloud TGU H645).
The five sizes of the circles
 correspond to different ranges of extinction.
Position of the eclipsing variable V375 Cep is shown by the cross.}
\label{6}
\end{figure}

\section{Discussion}

In the present investigation we find the following parameters of NGC
7142:  the mean extinction $A_V$ = 1.1 mag (corresponding to $E_{Y-V}$ =
0.28 or $E_{B-V}$ = 0.35), the distance 2.3 kpc (the true distance
modulus 11.8) and the age close to 3.0 Gyr.  The value of the distance
is not completely independent, since it is based on the five RCGs
identified by \citet{Sandquist2011} and confirmed in the present paper.
The parameters of the cluster are in a good agreement with the Sandquist
et al. results based on {\it BVI} and partly on {\it JHK} photometry.
However, in our case the results are obtained in a completely different
system which allowed us to classify stars in spectral and luminosity
classes and apply their individual dereddening.  Therefore the resulting
parameters of the cluster should be more reliable.

In most earlier investigations the parameters of NGC 7142 were compared
with the parameters of other old metal-rich clusters, mostly with M67.
According to the latest estimates \citep{Salaris2004, Cheng2012}, the
age of M67 is 4.3 Myr, i.e. it is somewhat larger than our value for NGC
7142.  Since the reddening of M67 is small and well known, and the NGC
7142 stars in our study are individually dereddened, we may directly
intercompare intrinsic colour indices of stars at turn-off points of
both clusters.  In our earlier study \citep{Boyle1998} it was found that
the turn-off point of M67 is at $(Y-V)_0$ = 0.50, this corresponds to
the spectral class close to F8, while in the present paper for NGC 7142
we have this colour at 0.46, corresponding to the spectral class F5.
This can be interpreted that M67 is really older because metallicities
of both clusters are not very different.

Another feature in the colour-magnitude diagram, depending on the age,
is the presence or absence of stars on the horizontal part of the
sequence of subgiants joining the hook above the turn-off point and the
lower part of the giant sequence.  This sequence in the clusters younger
than M67 is located in the lower part of the Hertzsprung gap and is
empty.  In NGC 7142 such stars are also absent, while in M67 the
subgiant sequence is well populated.  The stars on the subgiant sequence
of NGC 7142 could not be lost in the dereddening process since the
reddening lines in the CMD are approximately parallel to this sequence.

Recently, Sandquist et al.  (2013) estimated the age of the detached
eclipsing binary V375 Cep, a member of NGC 7142, using the masses and
radii of the components determined by modelling radial velocity and
light curves of the system.  The measured mass and radius of the primary
component gives an age of 3.3--3.6 Gyr.  The lower limit of this age is
not very different from our value of 3.0 Gyr, which is also not very
accurate due to the input physics used in computing the isochrones and
their transformation to the observational CMD with the theoretical
model atmospheres and synthetic spectral energy distributions.  The star
V375 Cep was not included in Table 2 since its color indices, determined
from our CCD frames in different filters, have been exposed in different
variability phases.  For the estimation of colour excess of V375 Cep, we
selected the exposures where the star is the brightest in the $Y$ and
$V$ passbands (both components are visible).  In these exposures, the
average values of $V$ and $Y$--$V$ are 16.090 and 0.739, respectively.

The combined intrinsic colour $(Y-V)_0$ of V375 Cep was calculated
in the following way.  The masses of the components in the
Sandquist et al. (2013) model correspond to the main-sequence stars of
spectral classes F4 and G9.  Taking the absolute magnitudes in the $Y$
and $V$ passbands for these spectral types, we have found the combined
intrinsic colour index of the binary, $(Y-V)_0$ = 0.45, which gives
$E_{Y-V}$ = 0.29 or $E_{B-V}$ = 0.36, i.e., the value which is very
close to the average colour excess of the cluster.  The star is plotted
in the intrinsic colour-magnitude diagram (Fig.\,4) with its
extinction-corrected magnitude at maximum brightness and the combined
intrinsic colour. Four other eclipsing variables from the
\citet{Sandquist2011} list, probably non-members, are also plotted.

\section{Results and conclusions}

1. Medium-band seven-colour photometry of 1037 stars in the
13$\arcmin$\,$\times$\,13$\arcmin$ area in the direction of the cluster
NGC 7142 in Cepheus is accomplished.

2. For 507 stars, using the interstellar reddening-free $Q$-parameters,
photometric spectral and luminosity classes in the MK system are
determined.  Down to $V$ = 18.5 mag, the amount of classified stars is
about 80\% of the observed stars.

3. The cluster distance, 2.30\,$\pm$\,0.16 kpc, is estimated using five
red clump giants from their photometric data in the Vilnius and 2MASS
systems.  Taking into account the errors of absolute magnitudes
$\pm$\,0.5 mag, possible members of the cluster in the range of
distances between 1.8 and 2.9 kpc are selected. More members close to
the red giant sequence were added from their membership estimations from
radial velocities. The total number of the possible cluster members is
255.

4. The possible cluster members, plotted in the dereddened
colour-magnitude diagram together with the Padova isochrones, are
applied to estimate the age of NGC 7142 which is found to be
3.0\,$\pm$\,0.5 Gyr.  In the diagram, five possible red clump giants,
two asymptotic branch stars and four blue stragglers are identified.

5. The interstellar extinction vs. distance diagram is plotted for all
stars of the field classified in spectral and luminosity classes.  The
mean extinction $A_V$ in the direction of the cluster is found to be
close to 1.1 mag corresponding to $E_{B-V}$\,$\approx$\,0.35.  The range
of variability of the extinction across the cluster area is about
$\pm$\,0.4 mag.

6.  The surface distribution of extinction in the area is quite
mottled but some small areas with the larger or smaller extinction than
the average can be identified.  The distribution of stars in the area is
filamentary, with a number of patches free of stars.

7. For the eclipsing variable V375 Cep, a cluster member, the
extinction is found to be $E_{Y-V}$ = 0.29 or $E_{B-V}$ = 0.36, which is
close to the average value for the cluster.

\section*{Acknowledgements} The use of the Simbad, WEBDA, ADS, SkyView
and Padova databases is acknowledged.  We are grateful to Eric L.
Sandquist for sending us his photometric catalogue, to L\'eo Girardi for
the calculation of isochrones in the Vilnius system and to the referee
for important comments.  The project is partly supported by the Research
Council of Lithuania, grant No.  MIP-061/2013.

\bibliographystyle{mn2e}
\bibliography{straizys}

\label{lastpage}

\end{document}